# Molecular gas observations and enhanced massive star formation efficiencies in M100

J. H. Knapen[1,2], J. E. Beckman[2], J. Cepa[2], and N. Nakai[3]

[1] Département de Physique, Université de Montréal, C.P. 6128, Succursale Centre Ville, Montréal (Québec), H3C 3J7 Canada; and Observatoire du Mont Mégantic. E-mail knapen@astro.umontreal.ca

[2] Instituto de Astrofísica de Canarias, E-38200 La Laguna, Tenerife, Spain

[3] Nobeyama Radio Observatory, Nobeyama, Minamimaki, Minamisaku, Nagano 384-13, Japan

**Abstract.** We present new $J = 1 \to 0$ $^{12}$CO observations along the northern spiral arm of the grand-design spiral galaxy M100 (NGC 4321), and study the distribution of molecular hydrogen as derived from these observations, comparing the new data with a set of data points on the southern arm published previously. We compare these measurements on both spiral arms and on the interarm regions with observations of the atomic and ionized hydrogen components.

We determine massive star formation efficiency parameters, defined as the ratio of H$\alpha$ luminosity to total gas mass, along the arms and compare the values to those in the interarm regions adjacent to the arms. We find that these parameters in the spiral arms are on average a factor of 3 higher than outside the arms, a clear indication of triggering of the star formation in the spiral arms. We discuss possible mechanisms for this triggering, and conclude that a density wave system is probably responsible for it. We discuss several possible systematical effects in some detail, and infer that the conclusions on triggering are sound. We specifically discuss the possible effects of extinction in H$\alpha$, or a non-standard CO to H$_2$ conversion factor ($X$), and find that our conclusions on the enhancement of the efficiencies in the arms are reinforced rather than weakened by these considerations.

A simple star forming scheme involving threshold densities for gravitational collapse is discussed for NGC 4321, and comparison is made with M51. We find that the gas between the arms is generally stable against gravitational collapse whereas the gas in the arms is not, possibly leading to the observed arm–interarm differences in efficiency, but also note that these results, unlike the others obtained in this paper, do depend critically on the assumed value for the conversion factor.

## 1. Introduction

Although it is clear that stars are forming at higher rates in spiral arms than in the rest of the discs of spiral galaxies, simple comparison of these rates does not tell too much about the physics of the massive star formation (SF) process. It is not clear whether the enhanced star formation rate (SFR) in the arms is due exclusively to and depends linearly on an enhanced neutral gas density there, or whether specific amplifying mechanisms are at work (see Elmegreen 1995 for a review). One can formulate a massive star formation efficiency (MSFE) parameter as the SFR per unit gas mass; or the H$\alpha$ luminosity, $L(\mathrm{H}\alpha)$, divided by the surface density of the total neutral gas, molecular plus atomic, $\sigma(\mathrm{H\,\scriptstyle I}+\mathrm{H}_2)$. Assuming the H$\alpha$ is emitted in ionization-bounded H II regions, $L(\mathrm{H}\alpha)$ is monotonically dependent on the number of ionizing photons, and thus the number of ionizing stars per unit area of galactic disc, while $\sigma(\mathrm{H\,\scriptstyle I}+\mathrm{H}_2)$ is a measure of the amount of neutral gas available for SF.

SF efficiency parameters in spiral discs defined in similar ways, have been studied by a number of authors (e.g. DeGioia-Eastwood et al. 1984; Thronson et al. 1989; Wiklind & Henkel 1989; Devereux & Young 1991), but for the problem stated above it is not satisfactory to determine global values for efficiencies. Tacconi & Young (1986) and Lord & Young (1990) discussed efficiencies in arm and interarm regions separately, but the resolution of their data was not high enough to sample the arms adequately. Cepa & Beckman (1990) and Knapen et al. (1992) did use ionized and neutral gas data at high enough resolution ($\sim 15''$) to cleanly map the arm regions and interarm regions separately. They defined the ratio, $\epsilon$, of the MSFE parameter in an arm, and in the neighbouring interarm disc. If the efficiencies in and outside the arms are the same, $\epsilon$ equals unity (this can of course be the case even when the SFR in the arms is enhanced, if the underlying gas density is similarly enhanced). A value of $\epsilon$ significantly higher than unity, however, implies directly that some mechanism amplifies, or triggers, the efficiency of massive SF in the arms as compared to the adjacent interarm region, in such a way that in the arms there is a non-linear dependence of the massive SFR on the underlying neutral gas density. Cepa & Beckman (1990) found high values for $\epsilon$ in the grand-design spirals NGC 3992 and NGC 628, implying triggering of massive SF in the spiral arms. They interpreted the strongly modulated and symmetrical behaviour of $\epsilon$ along the arms as signatures of a density-wave.

Knapen et al. (1992) applied the technique to M51, including the $H_2$ column density (which dominates within the inner 5 kpc) in the analysis. Results similar to those in the previous objects were found in M51, i.e. $\epsilon$ showed peaks ranging in some cases to well over 10 in amplitude, whose radial distances from the centre of M51 were essentially the same for both arms, as were the distances of the troughs between the peaks. Here again resonance structure in $\epsilon$, as well as triggering of massive SF by the dynamical effects within the resonance pattern, were picked out clearly.

Given the symmetry of the $H_2$ molecule, which causes its dipolar rotational transitions to be suppressed, measurement of CO emission is the only practically applicable method for determining the amount of molecular hydrogen in the disc of a galaxy. It is measured in the emission lines of the CO molecule (in this case the line corresponding to the $^{12}CO\ J = 1 \rightarrow 0$ transition). There are problems however with this method of determining $H_2$ parameters. One is that the line is almost always optically thick, the second that the amount of observed CO may not be monotonically related to an $H_2$ column density. We will discuss possible implications of these uncertainties in some more detail below, but will assume for the moment that we can convert the observed CO emission into a column density of molecular hydrogen using a single conversion factor (hereafter $X$), which is a satisfactory approximation based on direct separate measurements of CO and $H_2$ within the Galaxy. A review on the use of CO to measure extragalactic $H_2$ densities is given by Young & Scoville (1991).

In the present paper we add another galaxy to the as yet very limited number of objects that has been studied in detail with the arm/interarm MSFE technique, using new observational data on the ionized, atomic and molecular hydrogen for NGC 4321 (=M100). This is a grand-design late-type galaxy with a weak stellar bar (Pierce 1986; Knapen et al. 1993). Earlier CO observations of this galaxy were made by Young & Scoville (1982), Young, Scoville and Brady (1985; only the central position) and Kenney & Young (1988) using the FCRAO antenna with a resolution of $50''$. Recently, García-Burillo, Sempere & Combes (1994) used the 30m IRAM telescope to map a region along the minor axis of NGC 4321 in the $^{12}CO\ 1 \rightarrow 0$ and $2 \rightarrow 1$ transitions at $21''$ and $12''$ resolution, and used their observational results as input for numerical modeling of the galaxy, leading to an estimate of the pattern speed of the density wave pattern in the galaxy (Sempere et al. 1995). Rand (1995) obtained interferometric CO observations of the central region and the Southern spiral arm using BIMA. We present new $^{12}CO\ 1 \rightarrow 0$ observations at $15''$ resolution with the Nobeyama millimetre dish of a number of points *along* the two main spiral arms of NGC 4321, and of points in the neighbouring interarm region (see also Cepa et al. 1992).

In this Paper, we will denominate as the north arm the arm originating at the end of the bar east of the nucleus, passing north of the centre and continuing toward the west and later (but outside the range of our CO observations and thus of our efficiency study) the south-west. The south arm starts west of the nucleus at the end of the bar, passes the central region on the south, and continues toward the east and north-east. We assume a distance of 17.1 Mpc to NGC 4321 (Freedman et al. 1994), 1.0 kpc in the disc of the galaxy thus corresponds to $12''$.

We will first (Section 2) present the new CO observations, and briefly describe the H I and H$\alpha$ data used. The CO emission is discussed in some more detail in Section 3, and the combination of the CO data with the H$\alpha$ and H I data, including a determination of the efficiencies and their arm/interarm ratios, is described in Section 4. The efficiency method is discussed critically in Section 5, and in terms of density waves in Section 6. Section 7 discusses a simple scheme for the observed efficiencies based on gravitational instabilities. The main conclusions are briefly summarized in Section 8.

## 2. Observational Material

### 2.1. CO Observations

We observed NGC 4321 twice using the 45m telescope at the Nobeyama Radio Observatory (NRO) in Japan, from March $4^{th}$ to $8^{th}$, 1991 and from January $18^{th}$ to $22^{nd}$, 1992. During the first run we observed points along the southern arm, as described by Cepa et al. (1992). In the second run we concentrated on the north arm. We present those new data points here. Since apart from the receiver used, the observational procedure was the same in both runs, we refer the reader to Cepa et al. (1992) for further details. The receiver used for the new set of observations was a SIS with a 2048 channel acousto–optical spectrometer providing an individual channel width of 250 kHz and a frequency coverage of 250 MHz. We observed the $^{12}CO$ (J= 1 → 0) transition near 115 GHz. At this frequency, the beam size (FWHM) is 15″, or slightly smaller than a typical arm width in this galaxy. The main beam efficiency is 0.45. Pointing was checked every ∼1.5 hours using maser and point–like continuum sources. Atmospheric water vapour and wind speed were low during the observations. Final integration times ranged from 15 to 58 minutes, made up from individual 20 s long integrations, with ∼50% of the points having total integrations larger than 40 minutes. The individual spectra were co–added, and baseline subtracted by polynomial fitting. We determined total integrated antenna temperatures ($\int T_a^* \delta v$) and the noise therein (r.m.s. of line–free channels). The observed points, with our assigned reference numbers, are shown in Fig. 1, in an RA-dec grid centred on the nucleus of the galaxy. The baseline subtracted spectra are shown in Fig. 2.

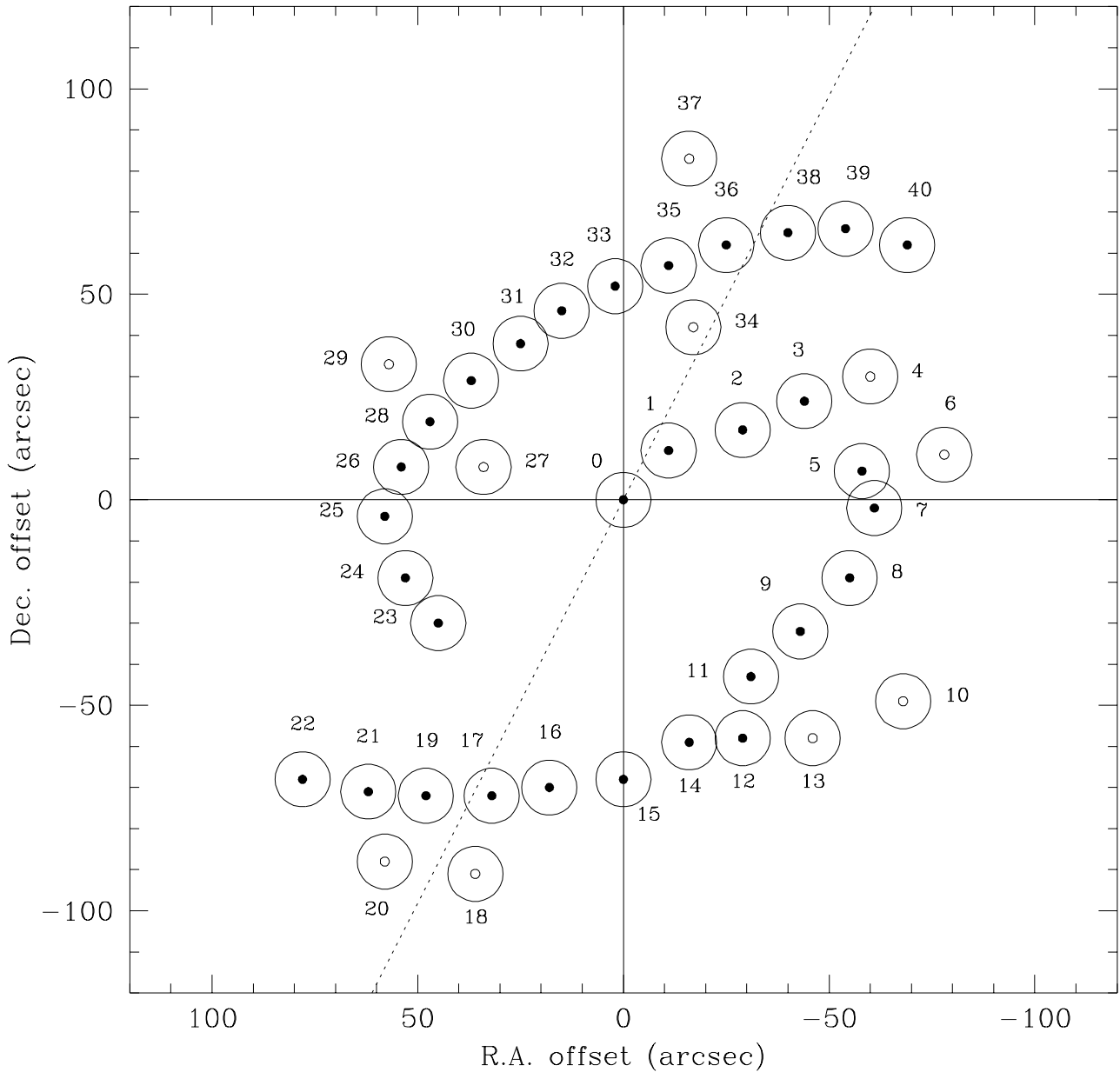

**Fig. 1.** Points in the disc of NGC 4321 observed with the 45 m NRO telescope, plotted on a grid indicating the distance in right ascension and declination from the centre of the galaxy, in arcsec (centre is at RA= $12^h\,22^m\,22^s\!.8$ (1950), $\delta = +16°\,06'\,00''$ (1950)). Large circles correspond to beam size (15″ FWHM). Filled symbols indicate arm points, open symbols interarm points. The dotted line is the position angle of the major axis, PA = 153°. Numbers assigned to observed points are indicated.

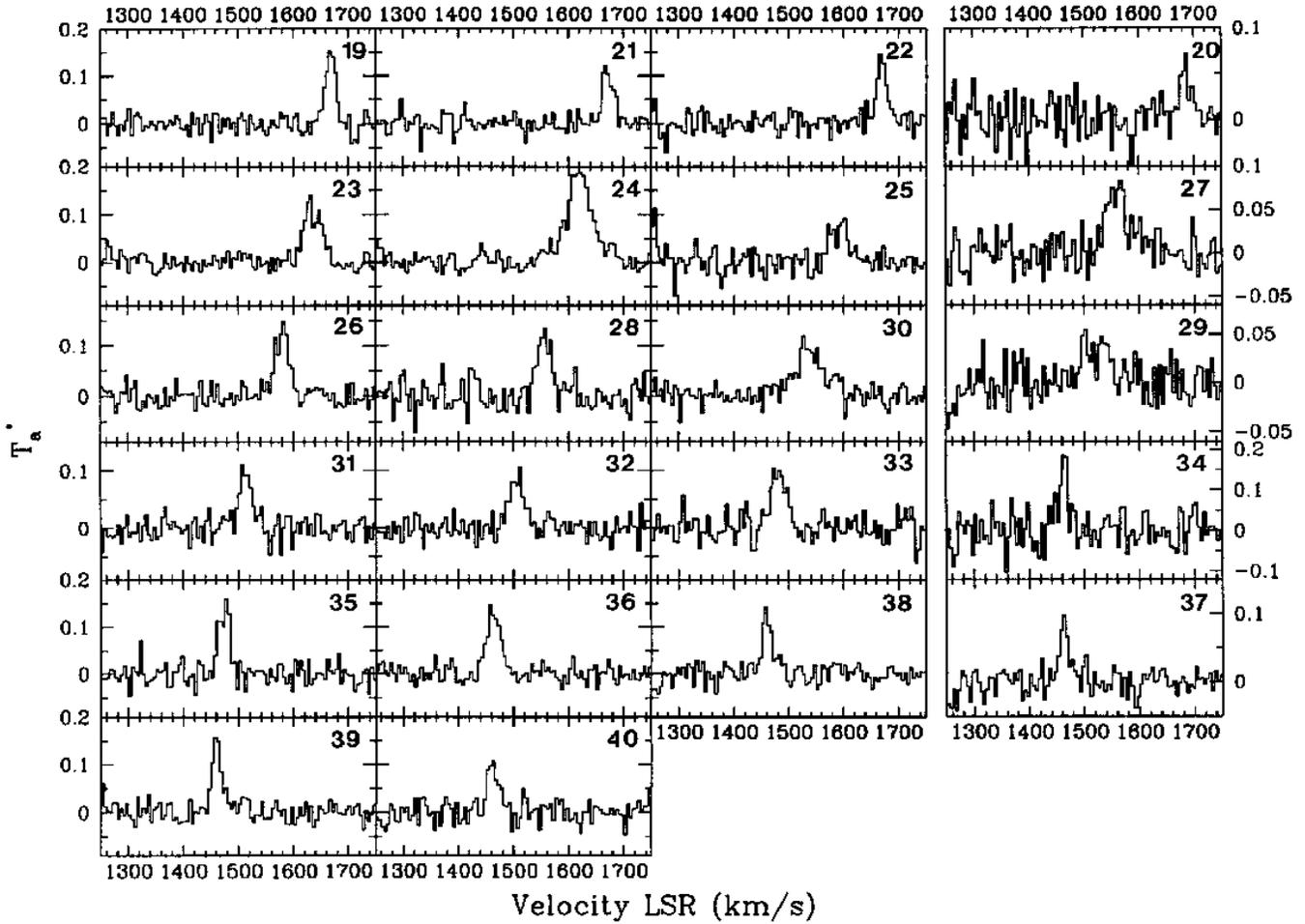

**Fig. 2.** Baseline-subtracted spectra of the observed points along the northern arm of NGC 4321, binned in intervals of 5 km/s. Antenna temperatures are not corrected for main beam efficiency (a factor of 2.2). Point numbers are indicated in the top right corner of each spectrum. Interarm points are in the right-hand column and their ordinate axes are to the right. Arm points are in the remaining columns and their ordinate axes are to the left.

In the rest of this paper (unless specifically indicated), we use a constant CO–H$_2$ conversion factor $X$ of $2.8 \times 10^{20}$ cm$^{-2}$ [K km s$^{-1}$]$^{-1}$ (Bloemen et al. 1986). The calibration (from $\int T_a^* \delta v$ to $N(H)$) is then

$$N(H) = 2.8 \cdot 10^{20} \times 2 \times \frac{1}{\eta_b} \times \int T_a^* \delta v,$$

with $N(H)$ in atoms cm$^{-2}$. The factor 2 in the calibration is required because the H$_2$ molecule consists of two hydrogen atoms, and $\eta_b$ is the main beam efficiency, $\eta_b = 0.45$.

The results are summarized in Table 1 (note that points 0–18 were published in Cepa et al. 1992). The first column is the number assigned to a given sky point. The distance to the centre $r$ (in arcsec) and the position angle $\theta$ on the plane of the sky (measured from N towards E) are given in columns 2 and 3. Column 4 is the deprojected distance to the centre $r_D$ in seconds of arc. We used an inclination angle $i = 27°$ (de Vaucouleurs et al. 1976) and a major axis position angle of 153° (Knapen et al. 1993) for the deprojection. Columns 5 and 6 are the offsets from the centre (at position RA= $12^h 22^m 22^s.8$ (1950), $\delta =$ +16° 06′ 00″ (1950)) in right ascension and declination, in seconds of arc (north and east are positive). Column 7 is the integrated antenna temperature $\int T_a^* \delta v$, and its error (the errors were evaluated from the line-free channels). $T_a^*$ is not corrected for the main beam efficiency. The last column (8) shows whether the points are measured on an arm or in the interarm disc; A means arm point, I interarm point, and C the centre of the galaxy.

### 2.2. Combination with Hα and H I data

The H I data used here were described in detail by Knapen et al. (1993). We use here the VLA total H I map at a resolution of 15″, comparable to the resolution of the CO data. The map units of mJy/B×m/s can be transformed into $N(H)$ by using the conversion

$$1 \text{ mJy/B} \times \text{m/s} = 4.89 \times 10^{18} \text{ atoms cm}^{-2}.$$

We used a new Hα continuum subtracted image of NGC 4321 obtained with the TAURUS camera in imaging mode on the 4.2m William Herschel Telescope. The observations and data reduction are described in detail by Knapen (1992) and Knapen (1996). In Fig. 3, we show a contour representation of the Hα map of NGC 4321. One instrumental count in the Hα image corresponds to $2.46 \times 10^{33}$ erg/s.

We now describe the additional steps followed to combine the CO, H I and Hα data. In the case of CO, we have only a limited number of data points, whereas in the case of H I and Hα, we can use a map covering the whole disc of NGC 4321. For a direct comparison of the three data sets, we are thus limited to the CO positions. The common spatial resolution used for all observations is ∼ 15″. Since the Hα map has a considerably higher intrinsic resolution (∼ 1″), we convolved the whole map to obtain a 15″ resolution image, using a gaussian "seeing" profile. In so doing we have degraded the resolution, and thereby the intensity of the Hα peaks and the arm/interarm contrast, to make the Hα directly comparable to the H I and CO data. Note that combining data sets of different resolutions will yield unphysical results.

The H I and Hα images, now both at 15″ resolution, were transformed to a common pixel grid. We chose to use 512 × 512 pixels of 1″ × 1″ for the combined grid, where the pixel sizes of the original maps were 4″ × 4″ in the case of the H I map, and 0″.27 × 0″.27 in the case of the Hα image. Using positions of field stars, we first gave the maps a common orientation and centre-point. Subsequently, the re-gridding was done using a bicubic interpolation algorithm. The Hα luminosity and the H I column density were determined directly from the re-gridded H I and Hα images. Average values over small boxes centred at the positions observed in CO were taken as the H I and Hα on-arm points. The rms of the > 100 pixels over which they were determined was taken as an indication of the error value in the points. A value for the interarm Hα or H I intensity for a specific point was determined as an average of measured intensities in the interarm regions measured at both sides of the arm. In most cases, values determined at opposite sides of the arms did not differ much. In the case of the second half of the north arm (the part toward the NW, or at position angle along the arm of > 60°), the determination of reliable interarm values was difficult, especially in H I. This is caused by the fact that the arm is not well defined there. In the case of Hα, it was still possible to determine the interarm emission for each point separately. In the case of H I though, we assumed a fixed interarm value for the complete second part of the arm, determined as an average value over that whole interarm region.

## 3. CO Emission

### 3.1. Radial behaviour

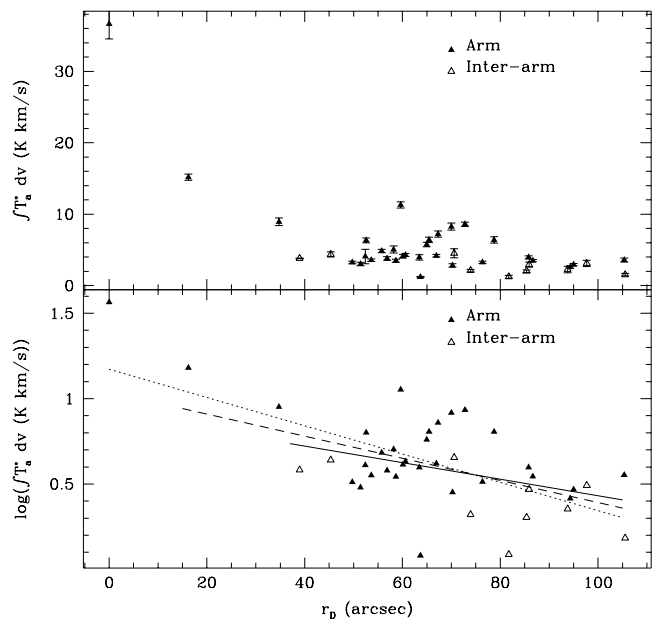

**Fig. 4. a.(top)** CO integrated intensity ($\int T_a^* \delta v$, in K km/s) as a function of deprojected distance to the centre $r_D$ of NGC 4321. Arm points are filled triangles, interarm points open triangles. **b. (bottom)** As Fig. 4a, now for log ($\int T_a^* \delta v$) as a function of $r_D$. Dotted, dashed and solid lines are exponential fits, with $h = 46″$, $h = 59″$ and $h = 76″$, respectively (see text).

In Fig. 4a we show the observed CO intensity as a function of deprojected radial distance for all measured points. In

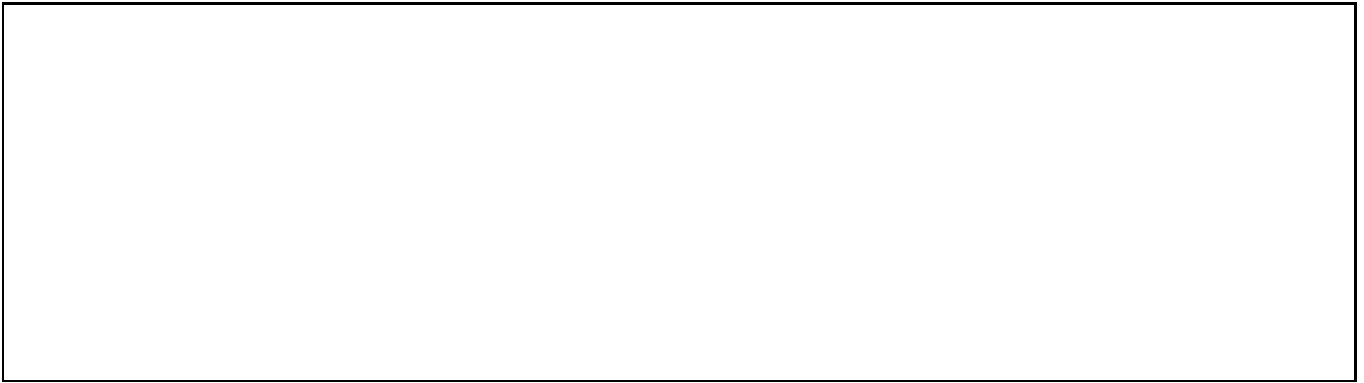

**Fig. 3.** Contour plot of the part of the continuum subtracted H$\alpha$ image of NGC 4321 used in the present paper, shown at the same scale and orientation as Fig. 1. Contour levels are 0.5, 1.5, 4.4, 13.3 and 26.6 $\times 10^{36}$ erg s$^{-1}$. The straight feature near RA= $+80''$, $\delta = -60''$ is an artifact in the image (probably a meteorite track).

Fig. 4b we plot the same points as in Fig. 4a, but now on an exponential scale, indicating exponential fits with straight lines here: the dotted line represents a fit with the central point included ($h_{\rm CO} = 44 \pm 8''$), the dashed line a fit where the centre was excluded ($h_{\rm CO} = 57 \pm 12''$), and the solid line a fit to all points excluding the central point and those in the bar ($h_{\rm CO} = 76 \pm 14''$). The values for the first two cases are only marginally different from the values given by Cepa et al. (1992) and Kenney & Young (1988). Young & Scoville (1982) determined a scale length of 58$''$ fitting their 3 disc points. Our value is also consistent with the minor axis CO profile shown by García-Burillo et al. (1994), although they do not explicitly fit an exponential to their data. Nakai (1992) described radial CO profiles in some other barred spirals, showing that the behaviour within and outside the bar region is different. Our fit to the points outside the bar indicates that indeed the profile is steeper when including the bar points. The presence of a distinct central component is clear, but we claim this is a (weak) bar component, and not just a central peak as claimed by Kenney & Young (1988) and Cepa et al. (1992).

### 3.2. CO intensity along the arms

In Fig. 5 we plot the integrated CO intensity against the position angle along the arms. The two main arms, and arm/interarm points are shown in different point styles. The arm points are connected by continuous lines, the interarm points by dashed lines. A position angle of 0° in the figure corresponds to the starting point of the two main spiral arms, defined here at the end of the bar (our CO data points 3 and 23). This zero-point corresponds to an angle of 120° (for the north arm), and 300° (south arm) on the galaxy (measured north over east). The position angle along the arm is measured clockwise.

It is immediately clear that the behaviour shown in Fig. 5 is quite different from the exponential fall-off seen in the previous section. The CO emission shows variations along the arms, but no generally falling trend is seen. This is mainly due to the fact that our arm points are concentrated in a small range of galactocentric radius. Although the sampling is limited, the interarm CO emission looks rather constant along the two arms. To estimate the interarm emission at a given point along an arm, determined here as the average between interarm emission from both sides of the arm point in question, it is im-portant to keep radial distance effects in mind. Along the north arm, we measured two pairs of two interarm points on each side of the spiral arm. In Fig. 5 it is seen that $\int T_a^* \delta v = 2.0 \pm 0.1$ K km/s for both the points at larger distances than the arm, while $\int T_a^* \delta v = 4.1 \pm 0.3$ K km/s for the points closer to the centre. We will use an average interarm CO emission of $\int T_a^* \delta v = 3 \pm 0.5$ K km/s for the whole north arm. Although the sampling is limited, this is justified by the symmetric distribution of the interarm points and by the fact that the emission from the first set of interarm points is almost equal to that from the second set. In the case of the south arm, all the interarm points measured lie somewhat further out in the disc than the arm points. We assume a general interarm value for the south arm of $\int T_a^* \delta v = 3 \pm 0.5$ K km/s, the same as that found for the north arm.

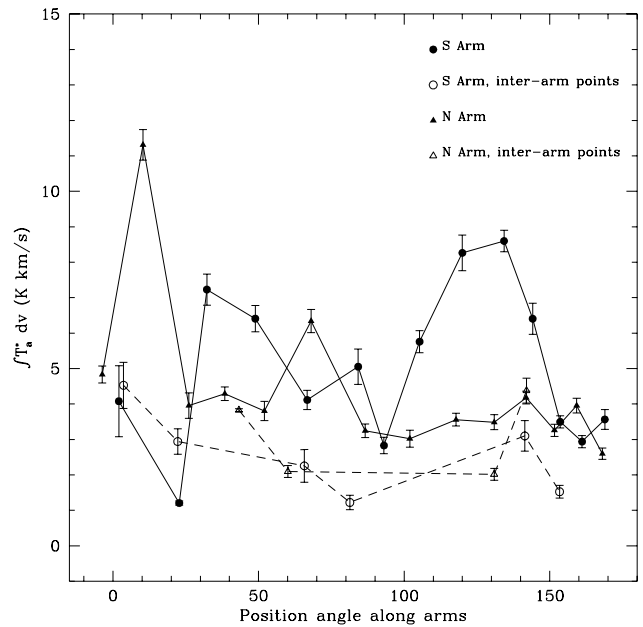

**Fig. 5.** CO integrated intensity ($\int T_a^* \delta v$, in K km/s) along the two main arms of NGC 4321. Filled symbols are arm points, open symbols are interarm points along the same arm. The points along the south arm are represented by circles, points along the north arm by triangles.

The profiles along the two arms show some pronounced differences. Along the south arm, the CO emission shows two broad peaks, centred at position angles of 30° and 120°, and reaching peak values of some 2−3 times the local interarm emission (Fig. 5). At either side of these peaks, the emission falls off to a level of $\int T_a^* \delta v = 3 - 4$ K km/s, equal to or only marginally higher than the interarm emission near those positions. The CO emission along the north arm looks quite different from that along the south arm. It is almost completely constant at a level of some $3 - 4$ K km/s, but shows two narrow peaks, near position angles 0° and 60°. Note however that both of these peaks consist of only one observed point.

The arm/interarm ratio varies between 1 and 3. This value is lower, but only a little lower, than the arm/interarm ratio in CO of $\sim 3$ as measured in M51 (see Knapen et al. 1992). We will discuss the arm and interarm CO profiles in more detail in the next Section, where we will combine the CO data with H I and H$\alpha$ observations.

## 4. Gas distribution and efficiencies along the arms

### 4.1. H$\alpha$, H I and H$_2$ along the arms

The arm and interarm gas densities and H$\alpha$ luminosities at the positions of the points observed in CO are shown in Fig. 6a for the south arm, and in Fig. 6b for the north arm, plotted as functions of position angle along the arms.

We will first discuss the south arm. The H$_2$ and the H$\alpha$ show a similar structure along this arm. A depression is seen between some 70° and 100° in position angle; further along the arm (higher PA) both $L_{H\alpha}$ and $\sigma_{H_2}$ are enhanced. In the first part of the arm, $L_{H\alpha}$ is slightly enhanced, and the $\sigma_{H_2}$ possibly as well. The atomic gas density is in general a factor 10 lower than the molecular gas density (this value depends on the exact value of $X$). Thus, as in the case of M51, the role of variations of $\sigma_{HI}$ in the MSFE will be limited. The $\sigma_{HI}$ along the arm shows hints of similar behaviour to the $\sigma_{H_2}$ and $L_{H\alpha}$, in particular it shows a depression near PA= 80° which coincides with the depression seen in the ionized and molecular gas tracers. There is also a depression in $\sigma_{HI}$ near PA= 20°, where we noticed the low CO emission discussed before. The interarm values vary little with increasing position angle along the arms. Arm/interarm contrasts are generally about 7 in H$\alpha$, about 2.5 in H I and about 2 in H$_2$.

The distribution of the gas tracers in the north arm (shown in Fig. 6b) is quite different from that in the south arm. The H$\alpha$ luminosity is seen to start at relatively high values when beginning at PA=0° along the arm. This is a region at the end of the bar where the SF is enhanced. Outside this region, $L_{H\alpha}$ falls off to low values, and stays low until about PA=150°. In this whole part, the north arm is poorly defined: the H$\alpha$ arm/interarm contrast is only between 1 and 3. At the end of the region observed in CO, the H$\alpha$ luminosity is larger. This is the beginning of a part of the arm west of the nucleus, which is strongly emitting in H$\alpha$. The H I density shows little structure along the north arm, and is generally at a lower level than the south arm. The H I arm/interarm contrast is only about 1.5 in the first part of the arm, and drops even further, to values close to unity, in the middle of the arm region observed in CO and considered here. $\sigma_{HI}$ seems to rise together with $L_{H\alpha}$ towards the end of the arm. As discussed before, the interarm emission is hard to determine in the outer part of the arm region considered in the present study. For those outer points we have assumed an average value over that whole part of the interarm region as a generic interarm value for $\sigma_{HI}$. The molecular hydrogen density is generally hardly enhanced with respect to the interarm value along the north arm. In general, we can remark that the north arm of NGC 4321 does not seem very interesting in terms of enhanced SF. It is difficult to define its form exactly due to the low contrast with the surrounding interarm regions. As can be seen from H I and particularly H$\alpha$ maps, the massive SF activity begins to rise where our CO observations had to stop.

### 4.2. MSFE arm/interarm ratios along the arms

Following Cepa & Beckman (1990) and Knapen et al. (1992), we define a massive star formation efficiency (MSFE) parameter as the H$\alpha$ luminosity divided by the total (atomic plus molecular) gas density at a certain point in a galaxy, which is closely proportional to the massive SFR per unit underlying gas mass. The arm/interarm ratio of the MSFE, $\epsilon$, is then defined as

$$\epsilon = \frac{\mathrm{MSFE}_A}{\mathrm{MSFE}_{IA}} = \frac{(L_{H\alpha})_A}{(L_{H\alpha})_{IA}} \times \frac{(\sigma_{HI} + \sigma_{H_2})_{IA}}{(\sigma_{HI} + \sigma_{H_2})_A}, \quad (1)$$

where the subscripts $A$ and $IA$ stand for arm and interarm, respectively. The values for $\epsilon$ along the two arms are plotted as a function of position angle along the two arms of NGC 4321 in Fig. 7.

The dashed line in the figure, at an $\epsilon$ of unity, indicates the case of equal MSFE's inside and outside an arm. This implies that if the massive SFR is higher in the arms, the underlying gas density is proportionally higher. As can be seen in Fig. 7, the values of $\epsilon$, as determined from the H$\alpha$, H I and H$_2$ data points described in the previous paragraph, are almost always significantly higher than unity, and in no case significantly lower. This indicates that there is more massive SF in the arms than might be expected from the amount of neutral gas available, as compared to the interarm regions. It thus means that the massive SF in the arms is enhanced, which implies that some specific triggering process is at work there.

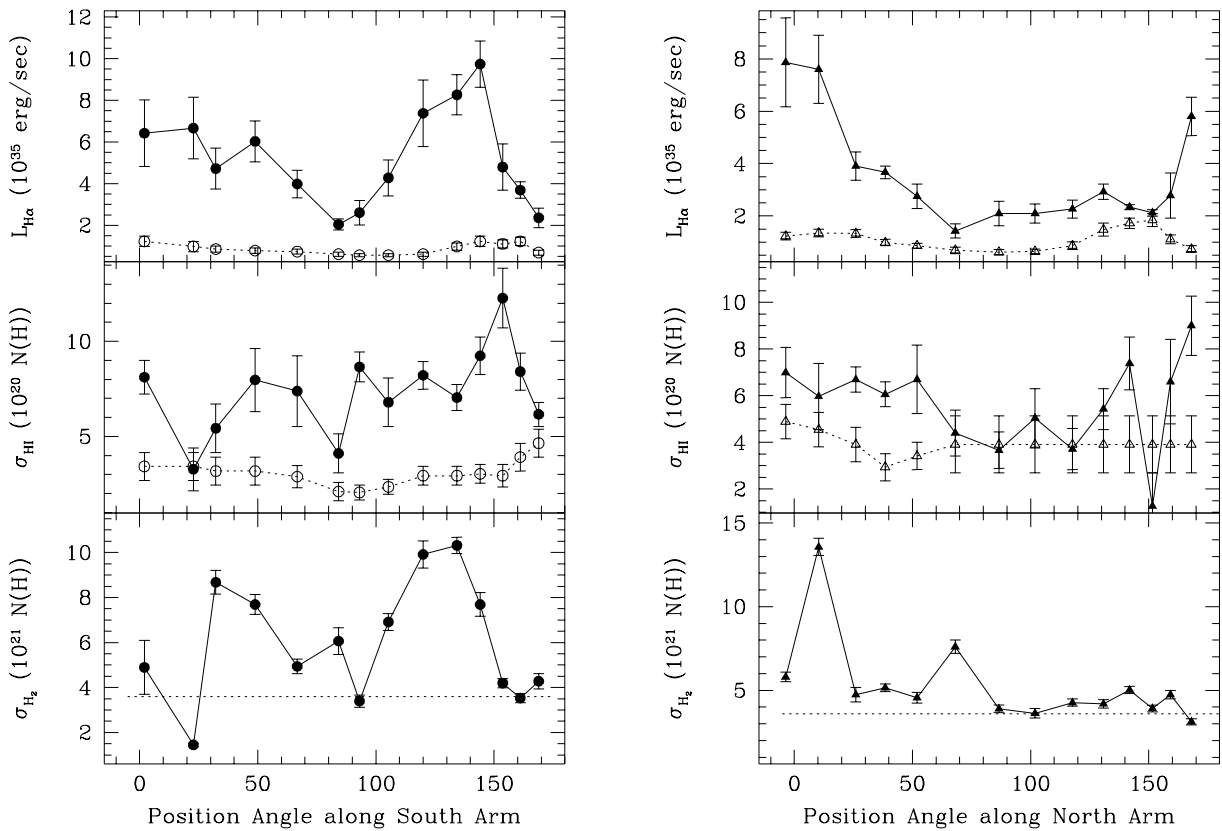

**Fig. 6. a.** H$\alpha$ (top panel), H I (middle panel) and H$_2$ (lower panel) distributions along the south arm of NGC 4321. For H$\alpha$ and H I, filled symbols are on-arm values, and open symbols indicate corresponding interarm values. The interarm value in H$_2$ is indicated by the dotted line in the lower panel. Note that values along the ordinate are an order of magnitude higher in H$_2$ than in H I. **b.** As Fig. 6a, now along the north arm of NGC 4321.

An average value for $\epsilon$ in the south arm is $\epsilon = 3.5 \pm 0.5$, whereas in the north arm it is somewhat lower, $\epsilon = 2.5 \pm 0.5$. It is important to discuss in some detail the peaks and dips seen in $\epsilon$ along the arms, since such peaks could originate in less reliable determinations of one of the gas tracers. An example of such a case may be the peak seen in the south arm at some 20° in position angle. This peak is not a result of enhanced massive SF in the arm, but rather of the low CO emission observed at that point (detection with low S/N). In the north arm, two separate points that were observed in CO showed significantly higher emission than their neighbour points. The result of the one-point CO peaks in the $\epsilon$ run along the north arm is a pair of relative dips, at positions angles of some 10° and 70° along the arm. Finally, there is the peak in $\epsilon$ seen at the eastern end of the north arm (at PA$\sim$ 170°). At this position, $L_{\mathrm{H}\alpha}$ is higher than at previous points, because one is starting to see the bright star forming complex west of the nucleus. Efficiencies may well be enhanced in this region, but more CO observations further along the arm are needed to confirm this.

Keeping these uncertainties in mind, the relations shown in Fig. 7 can be summarized as follows. In both arms, the values of $\epsilon$ are clearly and significantly higher than unity. Typical values for the arm/interarm contrast in MSFE in the part of the disc of NGC 4321 studied here are around 3. There is no convincing evidence for the existence of peaks and dips in $\epsilon$ along the arms of the high amplitudes seen in M51 (of order 10-20; Knapen et al. 1992), or even of lower amplitude. No symmetric pat-terns, and in fact no clear patterns at all, are seen in the runs of $\epsilon$ for the two different arms.

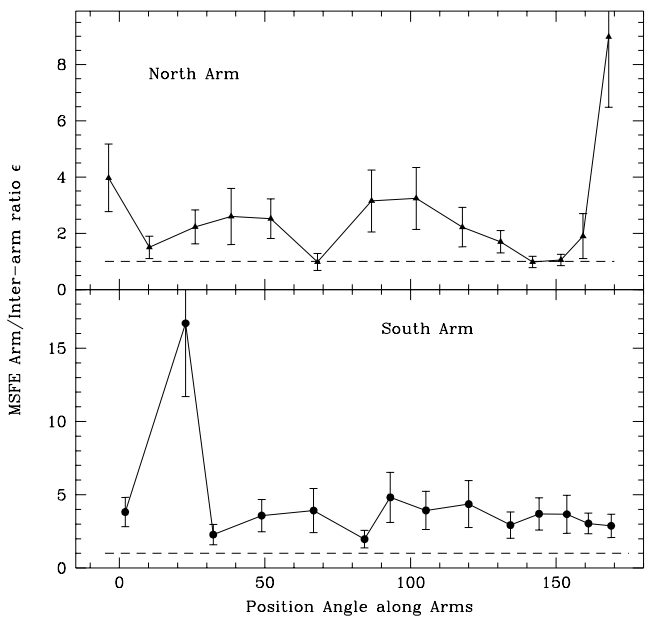

**Fig. 7.** Arm/interarm ratio of the MSFE ($\epsilon$) along the two main arms of NGC 4321: north arm in the upper panel, south arm in the lower panel. Dashed line at $\epsilon = 1$ indicates equal efficiencies in and outside the arms.

Equation (1) shows that high values of the arm/interarm ratio of the MSFE can in principle be caused by any of the four factors: H$\alpha$ luminosity and total gas density, in and outside the arms. Enhanced $L_{\mathrm{H}\alpha}$ on the arms or $\sigma_{\mathrm{gas}}$ between the arms, or reduced interarm $L_{\mathrm{H}\alpha}$ or on-arm $\sigma_{\mathrm{gas}}$ will produce higher $\epsilon$'s. A combination of more than one of these factors can reduce or amplify the effect. For NGC 4321, the profiles in Fig. 6 show that the enhanced efficiency in the arms is due only to an arm/interarm contrast in H$\alpha$ higher than in neutral gas (thus again to enhanced on-arm H$\alpha$ emission). Interarm H$\alpha$ emission or neutral gas densities do not "conspire" to suppress any H$\alpha$ peaks which might in principle be present, but which do not show up in the run of $\epsilon$ shown in Fig. 7. The values of $\epsilon$ larger than unity do indicate an enhanced MSFE in the spiral arms.

## 5. Discussion: on the validity of $\epsilon$ as an indicator of enhanced on-arm efficiencies

### 5.1. The CO to H$_2$ conversion factor $X$

The use of arm/interarm ratios in $\epsilon$ mitigates problems due to the need to convert observed CO column densities to H$_2$ column densities. While there is evidence that $X$ does not in fact vary by large factors across most of the disc of the Galaxy (Bloemen et al. 1986; Scoville & Good 1989), and that the Galactic value is valid in some external galaxies (Vogel, Boulanger & Ball 1987; Wilson & Scoville 1990), there are also indications that $X$ may vary within the Galaxy (e.g. Polk et al. 1988; Digel et al. 1995) or among galaxies (e.g. by as much as a factor of three in the well-studied galaxy M51: Adler et al. 1992; Rand 1993a; Nakai & Kuno 1995). There is however no evidence to date that $X$ is different in NGC 4321 (Rand 1995). There is no evidence for, nor against, arm-interarm variations in $X$ in general. The fact that $\epsilon$ as used in the present paper contains a ratio of $\sigma_{\mathrm{H}_2}(\mathrm{arm})/\sigma_{\mathrm{H}_2}(\mathrm{interarm})$, with both $\sigma$ at the same galactocentric radius, certainly means that a different value for $X$ for NGC 4321, or a radial gradient in $X$, will have no effect on our determination of $\epsilon$.

A potentially more fundamental problem with the use of CO as an indicator of the amount of molecular hydrogen has been exposed recently in a number of papers. From a study of molecular clouds in the SMC, Rubio, Lequeux & Boulanger (1993) concluded that $X$ is significantly different there from the "standard" Galactic value, and that the CO luminosity depends primarily on the UV radiation field and, to a lesser extent, on the cosmic-ray (CR) flux (see also Lequeux et al. 1994). Allen & Lequeux (1993) detected CO in the inner disk of M31. The UV radiation field and the CR flux are both very low in the regions where the "cold" CO was observed, and kinematic temperatures of the clouds can drop to less than 5 K (Allen et al. 1995). All this obviously implies that $X$ changes with the UV and CR flux, with in general relatively more CO luminosity per unit molecular hydrogen in zones with higher UV and CR flux. Note that $X$ may increase again in regions of very high UV flux, due to the higher degree of destruction of CO compared to that of H$_2$, and that also abundances may affect the CO luminosity.

Although the extragalactic observational evidence for these ideas is as yet limited to the SMC and M31, it is worthwhile to discuss their possible influence on the results of the present paper. Our main conclusion on triggering of the massive SF in the arm regions is based on our observations of an arm/interarm contrast that is larger in H$\alpha$ than in CO. We can expect that both the UV radiation field and the cosmic-ray flux are larger in the arms than between the arms, and thus the CO luminosity would be higher in the arms, following the reasoning laid out in the previous paragraph. In that case, $X$ would be higher in the interarm regions, and there would be more molecular hydrogen between the arms than one would assume using a constant $X$. But this means that the arm/interarm contrast in molecular hydrogen is *lower* than assumed, that $\epsilon$ is *higher*, and thus that the values for the arm/interarm efficiency as determined in Section 4 are in fact *lower limits* to the true values. This reinforces our conclusions on triggering of the massive SF in the arms.

The fact that in the outer parts of the disc of NGC 4321 (and also M51), where the neutral gas density is dropping below $10^{21}\,\mathrm{cm}^{-2}$, and where H I begins to compete with $H_2$ for dominance in column density, the behaviour of $\epsilon$ shows values consistently in the range close to 3, indicates that this value for $\epsilon$ cannot be an artifact of a wrongly adopted value for $X$. We feel it safe to conclude that our evidence for the triggering of the massive SF in the arms is sound.

## 5.2. Extinction by dust

Dust extinction could in principle affect our measurements of the H$\alpha$ emission in the arm and interarm regions, in the sense that one would expect more dust in the arms, and thus underestimate the arm H$\alpha$ emission compared to the interarm emission. If this were the case, the intrinsic arm/interarm ratio in H$\alpha$ would be larger than measured, and this means that the values for $\epsilon$ can only be larger than determined here. Beckman et al. (1995) studied in detail the dust content and distribution in NGC 4321 through the comparison of scale lengths in the optical, separately for arm and interarm regions. They found that there is not much dust in this galaxy, that there is significantly less dust between the arms than in them, and that the dust in the arms is mostly confined to narrow dust lanes. It thus seems safe to assume that the interarm H$\alpha$ is very little, if at all, affected by dust extinction. The H$\alpha$ emission in the arms arises mostly from regions outside the well-defined dust lanes, so the effect of extinction is also small in the arms. However, if we are in fact underestimating the on-arm H$\alpha$, e.g. because of emission arising inside the dust lanes which will thus be heavily obscured, we are also underestimating the arm-interarm contrast in H$\alpha$, and thus the efficiency ratios. The absence of peaks, as seen e.g. in M51 (Knapen et al. 1992), in H$\alpha$ along the arms of NGC 4321 cannot be caused by dust extinction, since this would imply the localized presence of huge quantities as dust which would show up in optical images, in the work of Beckman et al. (1995), and/or as strong neutral gas concentrations. There is no observational evidence for any of these effects, but in any case their presence would mean that our numbes for arm/interarm contrast in MSFE were lower limits and our qualitative conclusions would be strengthened, not weakened.

## 6. Efficiencies and density waves in NGC 4321

Comparing the results for the arm/interarm ratio of MSFE along the arms of NGC 4321 with those along the arms of M51 (Knapen et al. 1992), some important differences are obvious, but some common features must also be mentioned. Whereas in M51 we observed a strong symmetric pattern of high, narrow peaks and low dips (with values close to unity) in the run of $\epsilon$ along the arms, in NGC 4321 no pattern can be seen, and $\epsilon$ is practically constant along the arms. High peak values in $\epsilon$ of 10-20 as seen in M51 are not convincingly present in NGC 4321, where $\epsilon$ takes moderate values, with the exception of one individual point on each arm, where the CO intensities are lowest, and consequently the uncertainties highest. Also, $\epsilon$ does not drop down to unity in the case of NGC 4321 (except for a few points in the north arm), whereas it does in M51, between the peaks of high efficiency. On the other hand, an average of $\epsilon$ over the arms in M51 is similar to that determined for NGC 4321, of about 3 ($\pm 0.3$).

Thus in NGC 4321, as in M51, the MSFE is enhanced in the arms with respect to the interarm regions. We can conclude that some kind of triggering mechanism enhances the massive SF in the arms of NGC 4321 (otherwise $\epsilon$ would be $\sim 1$), but that the resonance pattern in $\epsilon$ is not apparent. Basically, what we have measured in NGC 4321 is that the arm/interarm ratio in $L_{H\alpha}$ is larger than that in $\sigma_{H_2}$, without any significant changes along the parts of the arms studied (note that, even though dependent on the exact $X$ value, the contribution of atomic hydrogen is of much reduced importance here, since it is far less abundant than molecular gas). In M51 (Knapen et al. 1992), and also in NGC 628 and NGC 3992 (Cepa & Beckman 1990), the enhanced efficiencies in the arms could be directly related to the presence of a density wave organizing the massive SF, through the symmetric pattern of peaks and dips in the run of $\epsilon$ along two main arms. The dips could in all these galaxies be identified with density wave resonances, and the global nature of this mechanism naturally explains the remarkable observed symmetry.

In NGC 4321, where no pattern in $\epsilon$ is seen in either arm, the efficiencies in the arms may still well be caused by density wave triggering. There is evidence from other observations that a density wave system is present in the disc of this galaxy, but also that its main resonances may be outside the region studied in the present paper. Streaming motions over the arms have been observed in H I (Knapen et al. 1993). Elmegreen, Elmegreen & Seiden (1989) were able to fit density wave resonances to radial intensity variations in arm profiles, as determined from computer enhanced optical images, and placed the co-rotation radius ($R_{CR}$) at $\sim 110''$ (well beyond the ends of the bar). This value for $R_{CR}$ has been confirmed by Sempere et al. (1995), who applied a two-dimensional test for reversal of streaming motions (following Canzian 1993) to an H I velocity field. However, $R_{CR}$ was estimated at around $75''$ by Arsenault et al. (1988), using an ionized gas rotation curve, and at $75'' \pm 7''$ by Canzian & Allen (1995), who optimized the fit of spiral density wave response functions from linear theory to an H$\alpha$ velocity field. This indicates a degree of confusion about where the co-rotation radius is located in this galaxy. It may be outside the bar region, and possibly outside the region studied in the present paper. Since a double inner Lindblad resonance is located well inside the region studied here (Arsenault et al. 1988; Knapen et al. 1995), we may be plotting $\epsilon$ between two resonances, and not covering the ILR or CR, which could explain why we do not see a clear pattern in the efficiencies.

## 7. Arm vs. Interarm Gravitational Instability

Galaxies with two symmetric spiral arms, such as NGC 4321 (where the general arm shape is symmetric, even though the distribution of SF regions in them is not) or M51, are believed to be controlled by spiral wavemodes (Bertin 1991), possibly organized by bars or companions. In the spiral arms, gas collapses into giant molecular complexes along an arm and may form OB stars (Scoville, Sanders & Clemens 1986). We see such a sweeping of gas in NGC 4321 as an enhancement of the neutral (molecular) gas density along the arms, as compared to the interarm regions. These clouds may lose enough of their supportive energy (turbulent, rotational or magnetic) or become massive enough to collapse under their own gravity, and may then form stars. Such triggering by gravitational instabilities is described in more detail by Elmegreen (1979, 1994) and Cowie (1981). Alternative mechanisms to increase the SF efficiency in the arms have been described (e.g. cloud-cloud collisions, Scoville et al. 1986). A review can be found in Elmegreen (1992).

Kennicutt (1989) studied massive SF in discs of spiral galaxies using information on the total gas content and on the massive SFR. He found that the massive SFR in the inner, dense regions can be described by a Schmidt (1959) power law $R = \alpha \rho_g^n$, with $n = 1.3 \pm 0.3$, where $R$, the massive SFR, depends only on the local gas density $\rho_g$. This relation breaks down in the outer parts of the galaxies Kennicutt studied, where the gas density drops below a critical threshold value. Massive SF is suppressed at densities well below this threshold, as the gas is stable against gravitational collapse at these densities.

An attractively simple explanation for the observed triggering in the arms of NGC 4321, and also in those of M51 (Knapen et al. 1992) might be that the gas density is higher than the local critical value in the arms, and falls below the threshold in the interarm regions. We can in fact check this hypothesis with the data at hand.

Critical densities $\sigma_c$ for several points along the arms and in the interarm regions can be derived following the formulae given by Kennicutt (1989). To make an estimate of $\sigma_c$, we assume that the rotation velocity is constant at $v_c = 220$ km s$^{-1}$ (from the H I rotation curve of Knapen et al. 1993). This is a reasonable approximation for all the points we consider, which are in a small range of radius around 70″. We then find that $\sigma_c = 3.2 \cdot 10^3 / x$ M$_\odot$ pc$^{-2}$, where $x$ is the distance from the centre, in arcsec. We determine total gas densities from the CO and H I observations (see above), add the densities in units of N(H), and multiply by 1.36 to account for the contribution of Helium to the total gas density, following Kennicutt (1989).

In Table 2, we show the gas densities and critical (threshold) densities for several arm and interarm points in NGC 4321. We express the results in solar masses per square parsec ($10^{21}$ N(H) cm$^{-2} \equiv 8.00$ M$_\odot$ pc$^{-2}$). In the south arm, the arm points have total gas densities about twice the critical density. For the interarm points $\sigma$ is just below the critical density in point 10, and slightly above it in point 18. These findings are in agreement with those of Rand (1995), who determined first the arm/interarm contrast in CO from streaming motions as observed with BIMA, and then surface and critical densities at arm and interarm points. In the north arm, which is, as we saw before, weaker than the south arm both in massive SF and gas density, the arm points have $\sigma$ higher than $\sigma_c$ for all the points but one, but only slightly higher. The interarm points though have gas densities that are below the critical densities by about a factor of two. The last point along the north arm, point no. 39, shows a greater contrast between $\sigma$ and $\sigma_c$. We saw before that at this point also $\epsilon$ is enhanced, thus the enhanced massive SF here may be related to the lower critical density.

We thus find that there is a significant difference between arm and interarm points, where gas densities of the arm points are either well above (south arm) or similar (less developed north arm) to the critical density. Interarm gas densities are very close (south) or well below (north arm) $\sigma_c$. In the arms the gas would thus not be stable against gravitational collapse, and in the interarm regions it might just be, which could well be the reason for the enhanced MSFE seen in the arms. This effect is more prominent along the south arm, which as we saw before shows substantially more SF than the north arm.

For M51, Rand (1993b) estimated the value of the epicyclic frequency $\kappa$ in the arm and in the interarm region, taking account of the streaming motions caused by the density wave. Rand (1993b) also used his results for $\kappa$ to estimate an arm/interarm contrast in CO, and combined those numbers with Lord's (1987) data to obtain a gas density. He found that whereas the arm gas is generally unstable against collapse ($\sigma_{H_2} > \sigma_c$), the interarm gas density is near the critical value. Oey & Kennicutt (1990) show the radial dependence of critical and observed gas densities for M51, using Lord's (1987) data for the CO.

As an extra test, we determine gas and critical densities for M51 by directly measuring the total gas densities from the H I (from Rots et al. 1990) and CO data (from Nakai et al. 1991) used by Knapen et al. (1992) at a few arm and interarm positions, and compare our results with those found by Oey & Kennicutt (1990) and Rand (1993b). The results for the three arm points (called A1, A2 and A3) and three interarm points (I1, I2 and I3) considered here are shown in Table 3. Points A3 and I3 are at a distance of 2.7 kpc in order to compare with Rand (1993b). Table 3 shows that the observed gas densities in both arm and interarm points lie well above the critical density. The fact that some of our points also lie well above the points determined by Oey & Kennicutt from Lord's data is probably an artifact of the lower resolution (45″) of Lord's observations, or of differences in the calibration of the data. Rand (1993b) listed critical and inferred surface densities at a distance of 2.7 kpc. For the arms he finds values for the gas density of 140 and 170 $M_\odot$ pc$^{-2}$ (not included in his numbers is the contribution of atomic gas, which is ~10%), which are comparable with our 196 $M_\odot$ pc$^{-2}$. Rand gives a value of $\sigma_c$=75-80 $M_\odot$ pc$^{-2}$ for the critical density, far lower than the measured density. For the interarm region, Rand (1993b) finds $\sigma = 45$ $M_\odot$ pc$^{-2}$ and $\sigma_c = 40$ $M_\odot$ pc$^{-2}$. Our value of $\sigma = 78$ $M_\odot$ pc$^{-2}$ for the interarm point at $R = 2.7$ kpc (which is a typical point for the interarm region at that galactocentric distance) is clearly higher. The difference may again be due to the lower resolution or different calibration of Lord's (1987) observations that Rand (1993b) used, or to the method Rand used to derive the arm/interarm contrast.

Explicitly considering the impact of changing $X$, we use a value of $N(H_2)/I_{CO} = 1.1 \times 10^{20}$ cm$^{-2}$ [K km s$^{-1}$]$^{-1}$, determined by Nakai & Kuno (1995) from a comparison of visual extinctions and CO data toward 30 H II regions in M51, a result which is in general agreement with conclusions from studies of virial and CO masses by Adler et al. (1992) and Rand (1993a). The observed and critical densities are much closer using the low value of $X$. Two of the arm points (A1 and A2) continue to have $\sigma$ $2-3$ times larger than $\sigma_c$, with A3 having a gas density only marginally larger than the local $\sigma_c$, while gas densities at the interarm points are all very close to the local critical densities, with $\sigma$ somewhat larger than $\sigma_c$ for I1, and somewhat smaller for I3. This is reminiscent of what we found for NGC 4321, where the arm gas may be unstable, but the interarm gas stable against gravitational collapse.

It is clear that any conclusion which is directly dependent on *absolute* measurements of quantities of molecular gas in external galaxies, such as those described in this section, but also e.g. the study of Kennicutt (1989), depend critically on the adoption of a correct for $X$. As we saw before (Sect. 5.1) there are indications that the standard conversion factor may not be correct in general. In M51, there is now mounting evidence that $X$ is in fact lower, as taken into account here. We note again, however, that so far there are no indications that $X$ is different from the "standard" value in NGC 4321 (see Rand 1995), but also that our conclusions on triggering (Sect. 4) seem robust against variations in $X$, unlike the results on gas vs. critical densities.

## 8. Conclusions

We have measured the $^{12}$CO J= 1 → 0 emission from a sample of points along the two main spiral arms of NGC 4321, and from a number of interarm points, using the Nobeyama 45 m millimetre wave telescope. In this paper we complement the data points presented by Cepa et al. (1992) for the south arm with new measurements along the north arm. The exponential decline of the radial distribution of the CO integrated antenna temperature, reported before, is confirmed, but the inner part of the radial profile is determined by emission from the weak bar of the galaxy. Interarm emission is constant at a level of $\int T_a^* \delta v = 3 \pm 0.5$ K km/s along the parts of the arms studied here. The south arm shows generally higher emission than the north arm.

Combining the CO data with new H I and H$\alpha$ maps, we have determined massive star formation efficiencies (defined as H$\alpha$ luminosity per unit neutral gas mass) *along* the two main spiral arms, and especially the arm/interarm ratio $\epsilon$ of these efficiencies. This is only the second galaxy, after M51 (Knapen et al. 1992) where $\epsilon$ has been determined along the spiral arms, including both the molecular and atomic hydrogen gas components in the analysis. As in M51, we find in NGC 4321 that $\epsilon$ is larger than unity, with an average value of $3 \pm 0.5$, very similar to the average value for M51. This implies that some mechanism *triggers* the massive star formation in the arms. Apart from a few localized peaks, values of $\epsilon$ are almost constant along both arms, and somewhat higher in the south than in the north arm. A density wave system is a good candidate for the triggering agent, even though direct evidence such as a symmetric pattern in $\epsilon$ similar to that found in M51 (Knapen et al. 1992) is absent in NGC 4321.

We find that in NGC 4321, measured gas densities are equal or larger than local critical densities for gravitational collapse in the arms, whereas the interarm gas densities are equal or lower than $\sigma_c$. We confirm earlier findings that in M51 both arm and interarm gas densities are higher than local critical densities, assuming the standard $X$-value. But assuming the lower value for $X$ reported in the literature for M51, a picture emerges that is very much like the one we find in NGC 4321. Thus in both galaxies gravitational instability could be a mechanism involved in causing the excess of massive SF in the arms, as compared to the interarm regions.

Our conclusion on triggering of the massive SF in the arms seems sound. It basically reflects a higher arm/interarm contrast in H$\alpha$ than in molecular gas, which is dominant by an order of magnitude over atomic gas in the inner region of the galaxy studied here. From the profiles along the arms for the several gas tracers, we see no evidence that the enhanced on-arm efficiencies are due to any effect other than enhanced on-arm massive SF. Dust extinction or changes in the CO to H$_2$ conversion factor ($X$) are among the effects we have eliminated as possible causes for enhanced efficiency in the arms. Both effects, if at all present, would *increase* the arm/interarm efficiency ratios and thus reinforce rather than limit our conclusions on triggering.

*Acknowledgements.* The 45-m telescope is operated by the Nobeyama Radio Observatory which is a branch of the National Astronomical Observatory, the Ministry of Education, Science, and Culture, Japan. The William Herschel Telescope is operated on the island of La Palma by the Royal Greenwich Observatory in the Spanish Observatorio del Roque de los Muchachos of the Instituto de Astrofísica de Canarias. JHK and JEB are partially supported by the Spanish DGICYT (Dirección General de Investigación Científica y Técnica) grant no. PB91-0510.

# References

Adler, D.S., Lo, K.Y., Wright, M.C.H., Rydbeck, G., Plante, R.L., Allen, R.J. 1992, ApJ 392, 497

Allen, R.J., Lequeux, J. 1993, ApJ 410, L15

Allen, R.J., Le Bourlot, J., Lequeux, J., Pineau des Forêts, G., Roueff, E. 1995, ApJ 444, 157

Arsenault, R., Boulesteix, J., Georgelin, Y., Roy, J.-R. 1988, A&A 200, 29

Beckman, J.E., Peletier, R.F., Knapen, J.H., Maté, M.J., Gentet, L. 1995, In: Opacity of Spiral Discs, Proc. NATO Adv. Workshop., Eds. J. Davies and D. Burstein, Kluwer, Dordrecht, p. 197

Bertin, G. 1991, in Dynamics of Galaxies and Their Molecular Cloud Distributions, Eds. F. Combes and F. Casoli, Kluwer Dordrecht, p.93

Bloemen, J.B.G.M. et al. 1986, A&A 154, 25

Canzian, B. 1993, ApJ 414, 487

Canzian, B., Allen, R.J. 1995, in preparation

Cepa, J., Beckman, J. E. 1990, ApJ 349, 497

Cepa, J., Beckman, J.E., Knapen, J.H., Nakai, N., Kuno, N. 1992, AJ 103, 429

Cowie, L.L. 1981, ApJ 245, 66

DeGioia-Eastwood, K., Grasdalen, G.L., Strom, S.E., Strom, K.M. 1984, ApJ 278, 564

de Vaucouleurs, G., de Vaucouleurs, A., Corwin, H.G.Jr. 1976, Second Reference Catalogue of Bright Galaxies (RC2), Univ. of Texas Press, Austin

Devereux, N.A., Young, J.S. 1991, ApJ 371, 515

Digel, S.W., Grenier, I.A., Heithausen, A., Hunter, S.D., Thaddeus, P. 1995, ApJ (submitted)

Elmegreen, B.G. 1979, ApJ 231, 372

Elmegreen, B.G. 1992, in: Star Formation in Stellar Systems, Ed. G. Tenorio-Tagle, M. Prieto, F. Sánchez, University Press, Cambridge, p. 381

Elmegreen, B.G. 1994, ApJ 425, L73

Elmegreen, B.G. 1995, in *The formation of the Milky Way*, Eds. E. Alfaro & G. Tenorio-Tagle, Cambridge Univ. Press, Cambridge, in press

Elmegreen, B.G., Elmegreen, D.M., Seiden, P.E. 1989, ApJ 343, 602

Freedman, W. et al. 1994, Nat 371, 757

García-Burillo, S., Sempere, M.J., Combes, F. 1994, A&A 287, 419

Kenney, J.D., Young, J.S. 1988, ApJS 66, 261

Kennicutt, R.C. 1989, ApJ 344, 685

Knapen, J.H. 1992, PhD thesis, Univ. of La Laguna

Knapen, J.H. 1996, in preparation

Knapen, J.H., Beckman, J.E., Cepa, J., van der Hulst, J.M., Rand, R.J. 1992, ApJL 385, L37

Knapen, J.H., Cepa, J., Beckman, J.E., del Rio, M.S., Pedlar, A. 1993, ApJ 416, 563

Knapen, J.H., Beckman, J.E., Shlosman, I., Peletier, R.F., Heller, C.H., de Jong, R.S. 1995, ApJ, 443, L73

Lequeux, J., Le Bourlot, J., Pineau des Forêts, G., Roueff, E., Boulanger, F., Rubio, M. 1994, A&A 292, 371

Lord, S.D. 1987, PhD. Thesis, Univ. of Massachusetts

Lord, S.D., Young, J.S. 1990, ApJ 356, 135

Nakai, N. 1992, PASJ 44, L27

Nakai, N., Kuno, N., Handa, T., Sofue, Y. 1991, in Dynamics of Galaxies and Their Molecular Cloud Distributions, Eds. F. Combes and F. Casoli, Kluwer Dordrecht, p.63

Nakai, N., Kuno, N. 1995, PASJ (submitted)

Oey, M.S., Kennicutt, R.C. 1990, in: The Interstellar Medium in External Galaxies: Summaries of Contributed Papers, Eds. D.J. Hollenbach and H.A. Thronson (NASA Conf. Publ. 3084), 309

Pierce, M.J. 1986, AJ 92, 285

Polk, K.S., Knapp, G.R., Stark, A.A., Wilson, R.W. 1988, ApJ 332, 432

Rand, R.J. 1993a, ApJ 404, 593

Rand, R.J. 1993b, ApJ 410, 68

Rand, R.J. 1995, AJ 109, 2444

Rots, A.H., Bosma, A., van der Hulst, J.M., Athanassoula, E., Crane, P.C. 1990, AJ 100, 387

Rubio, M., Lequeux, J., Boulanger, F. 1993, A&A 271, 9

Schmidt, M. 1959, ApJ 129, 243

Sempere, M.J., García-Burillo, S., Combes, F., Knapen, J.H. 1995, A&A 296, 45

Scoville, N.Z., Good, J. 1989, ApJ 339, 149

Scoville, N.Z., Sanders, D.B., Clemens, D.P. 1986, ApJ 310, L77

Tacconi, L.J., Young, J.S. 1986, ApJ 308, 600

Thronson, H.A., Tacconi, L., Kenney, J., Greenhouse, M.A., Margulis, M., Tacconi-Garman, L., Young, J.S. 1989, ApJ 344, 747

Vogel, S., Boulanger, F., Ball, R. 1987, ApJ 321, L145

Wiklind, T., Henkel, C. 1989, A&A 225, 1

Wilson, C., Scoville, N.Z. 1990, ApJ 363, 435

Young, J.S., Scoville, N.Z. 1982, ApJ 258, 467

Young, J.S., Scoville, N.Z. 1991, ARA&A, 29, 581

Young, J.S., Scoville, N.Z., Brady, E. 1985, ApJ, 288, 487

**Table 1.** Summary of CO Observations. Columns are (respectively) point number, galactocentric distance, position angle and deprojected distance from the centre of the galaxy, RA and dec offsets from the centre of the galaxy, CO integrated flux and error, and arm (A), interarm (I) or centre (C) point identification.

| N | $r$ ($''$) | $\Theta$ (°) | $r_D$ ($''$) | RA ($''$) | Dec ($''$) | $\int T_a^* \delta v$ (K km s$^{-1}$) | A/I/A |
|---|---|---|---|---|---|---|---|
| 0 | 0.0 | 0.0 | 0.0 | 0 | 0 | 36.7±2.1 | C |
| 1 | 16.1 | 317.8 | 16.2 | -11 | 12 | 15.2±0.4 | A |
| 2 | 33.6 | 300.9 | 34.8 | -29 | 17 | 8.9±0.5 | A |
| 3 | 50.3 | 298.0 | 52.4 | -44 | 24 | 4.1±1.0 | A |
| 4 | 67.5 | 296.4 | 70.6 | -60 | 30 | 4.5±0.7 | I |
| 5 | 58.7 | 277.3 | 63.7 | -58 | 7 | 1.2±0.1 | A |
| 6 | 79.2 | 277.8 | 85.9 | -78 | 11 | 2.9±0.4 | I |
| 7 | 61.1 | 267.8 | 67.3 | -61 | -2 | 7.2±0.4 | A |
| 8 | 58.4 | 251.1 | 65.5 | -55 | -19 | 6.4±0.4 | A |
| 9 | 53.7 | 233.3 | 60.1 | -43 | -32 | 4.1±0.3 | A |
| 10 | 83.8 | 234.3 | 93.8 | -68 | -49 | 2.3±0.5 | I |
| 11 | 53.0 | 215.8 | 58.2 | -31 | -43 | 5.1±0.5 | A |
| 12 | 64.9 | 206.9 | 70.2 | -29 | -58 | 2.8±0.2 | A |
| 13 | 74.1 | 218.7 | 81.7 | -46 | -58 | 1.2±0.2 | I |
| 14 | 61.5 | 194.7 | 65.0 | -16 | -59 | 5.8±0.3 | A |
| 15 | 68.2 | 180.0 | 70.0 | 0 | -68 | 8.3±0.5 | A |
| 16 | 72.4 | 165.7 | 72.8 | 18 | -70 | 8.6±0.3 | A |
| 17 | 78.7 | 155.9 | 78.7 | 32 | -72 | 6.4±0.4 | A |
| 18 | 97.6 | 158.5 | 97.7 | 36 | -91 | 3.1±0.4 | I |
| 19 | 86.5 | 146.3 | 86.7 | 48 | -72 | 3.5±0.2 | A |
| 20 | 105.4 | 146.6 | 105.6 | 58 | -88 | 1.5±0.2 | I |
| 21 | 94.3 | 138.9 | 95.0 | 62 | -71 | 2.9±0.2 | A |
| 22 | 103.5 | 131.1 | 105.3 | 78 | -68 | 3.6±0.3 | A |
| 23 | 54.1 | 123.7 | 55.7 | 45 | -30 | 4.8±0.2 | A |
| 24 | 56.3 | 109.7 | 59.6 | 53 | -19 | 11.3±0.4 | A |
| 25 | 58.1 | 94.0 | 63.5 | 58 | -4 | 4.0±0.4 | A |
| 26 | 54.6 | 81.6 | 60.6 | 54 | 8 | 4.3±0.2 | A |
| 27 | 34.9 | 76.8 | 39.0 | 34 | 8 | 3.8±0.0 | I |
| 28 | 50.7 | 68.0 | 56.9 | 47 | 19 | 3.8±0.3 | A |
| 29 | 65.9 | 59.9 | 73.9 | 57 | 33 | 2.1±0.2 | I |
| 30 | 47.0 | 51.9 | 52.6 | 37 | 29 | 6.3±0.3 | A |
| 31 | 45.5 | 33.3 | 49.8 | 25 | 38 | 3.2±0.2 | A |
| 32 | 48.4 | 18.1 | 51.4 | 15 | 46 | 3.0±0.2 | A |
| 33 | 52.0 | 2.2 | 53.6 | 2 | 52 | 3.6±0.2 | A |
| 34 | 45.3 | 338.0 | 45.4 | -17 | 42 | 4.4±0.4 | I |
| 35 | 58.1 | 349.1 | 58.6 | -11 | 57 | 3.5±0.2 | A |
| 36 | 66.9 | 338.0 | 66.9 | -25 | 62 | 4.2±0.2 | A |
| 37 | 84.5 | 349.1 | 85.4 | -16 | 83 | 2.0±0.2 | I |
| 38 | 76.3 | 328.4 | 76.4 | -40 | 65 | 3.3±0.2 | A |
| 39 | 85.3 | 320.7 | 85.8 | -54 | 66 | 4.0±0.2 | A |
| 40 | 92.8 | 311.9 | 94.3 | -69 | 62 | 2.6±0.2 | A |

**Table 2.** Surface and critical gas densities in NGC 4321. Column 1: point identification as in Table 1 and Fig. 1. Column 2: point is in south (S) or north (N) arm, and (column 3) point is in the arm (A) or in the interarm (I) region. Column 4: deprojected distance from the centre, in arcsec. Column 5: total gas density, in M$_\odot$ pc$^{-2}$. Column 6: critical or threshold gas density, in M$_\odot$ pc$^{-2}$.

| Point | Arm S/N | A/I | $r_D$ ($''$) | $\sigma$ (M$_\odot$ pc$^{-2}$) | $\sigma_c$ (M$_\odot$ pc$^{-2}$) |
|---|---|---|---|---|---|
| 8 | S | A | 66 | 90 | 49 |
| 15 | S | A | 70 | 111 | 46 |
| 10 | S | I | 94 | 38 | 34 |
| 18 | S | I | 98 | 41 | 33 |
| 26 | N | A | 61 | 63 | 52 |
| 32 | N | A | 51 | 49 | 63 |
| 36 | N | A | 67 | 58 | 48 |
| 39 | N | A | 86 | 61 | 37 |
| 27 | N | I | 39 | 48 | 82 |
| 29 | N | I | 74 | 26 | 43 |

**Table 3.** Critical densities $\sigma_c$ as determined by Oey & Kennicutt (1990) and Rand (1993b; for points at 2.7 kpc from the centre) for a small number of typical points in M51 (column 4). Gas densities $\sigma$ as determined from the CO and H I maps, for different values of $X$: the standard value of $2.8 \times 10^{20}$ (column 5), and $1.1 \times 10^{20}$ (column 6). Points A1, A2 and A3 are on the arms, I1, I2 and I3 are in the interarm regions.

| Point | Distance ($''$) | (kpc) | $\sigma_c$ (M$_\odot$ pc$^{-2}$) | $\sigma_{X=2.8}$ (M$_\odot$ pc$^{-2}$) | $\sigma_{X=1.1}$ (M$_\odot$ pc$^{-2}$) |
|---|---|---|---|---|---|
| A1 | 110 | 5.1 | 20 | 88 | 35 |
| A2 | 110 | 5.1 | 20 | 141 | 55 |
| A3 | 58 | 2.7 | 80 | 196 | 77 |
| I1 | 43 | 2.0 | 45 | 136 | 53 |
| I2 | 48 | 2.2 | 45 | 112 | 44 |
| I3 | 58 | 2.7 | 40 | 78 | 31 |